\documentclass[twocolumn,aps,prl,superscriptaddress,preprintnumbers,nofootinbib]{revtex4-1}
\usepackage{graphicx}
\usepackage{subcaption}
\usepackage{color}
\graphicspath{{figures/}{fig/}}
\usepackage{amsmath}
\usepackage{amssymb}
\usepackage{bm}
\usepackage{slashed}
\usepackage{epsfig}
\usepackage{amsfonts}
\usepackage{epstopdf}
\usepackage{hyperref}
\usepackage{bbm}
\usepackage{textcomp}
\usepackage{color}
\newcommand{\sect}[1]{{\it \textbf{#1.} --- }}
\newcommand{\NOdisplay}[1]{ }

\def\MSbar{\overline{\mathrm{MS}}}
\def\als{\alpha_s}

\newcommand{\Oals}[1]{\mathcal{O}(\alpha_s^{#1})}

\def\mbos{m_{b,\mathrm{os}}}
\def\mbkin{m_{b,\mathrm{kin}}}

\def\mcOs{m_{c,\mathrm{1s}}}

\def\HTensor{\mathcal{W}_{Qq}^{\,\mu \nu}}

\def\BXulv{B \rightarrow X_u \ell \bar{\nu}_{\ell}}
\def\bulv{b \rightarrow u \ell \bar{\nu}_{\ell}}
\def\GamBulv{\Gamma^{\mathrm{sl}}_{b\rightarrow u}}
\def\Bpilv{B \rightarrow \pi \ell \bar{\nu}_{\ell}}

\def\cqlv{c \rightarrow q\,\ell \bar{\nu}_{\ell}}
\def\Dqlv{D \rightarrow X\,\ell \bar{\nu}_{\ell}}

\def\Gamsl{\Gamma^{\mathrm{sl}}}

\def\BXclv{B \rightarrow X_c \ell \bar{\nu}_{\ell}}

\begin{document}
\preprint{CPTNP-2026-006, SLAC-PUB-260209-1, ZU-TH 06/26}

\title{
Triple Differential Heavy-to-light Semi-leptonic Decays at Next-to-Next-to-Next-to-Leading Order in QCD
}

\author{Long Chen}
\email{longchen@sdu.edu.cn}
\affiliation{School of Physics, Shandong University, Jinan 250100, China}

\author{Xiang Chen}
\email{xiang.chen@physik.uzh.ch}
\affiliation{Physik-Institut, Universit\"at Z\"urich, Winterthurerstrasse 190, CH-8057 Z\"urich, Switzerland}

\author{Xin Guan}
\email{guanxin@slac.stanford.edu}
\affiliation{SLAC National Accelerator Laboratory, Stanford University, Stanford, California 94039, USA}

\author{Yan-Qing Ma}
\email{yqma@pku.edu.cn}
\affiliation{School of Physics, Peking University, Beijing 100871, China}
\affiliation{Center for High Energy Physics, Peking University, Beijing 100871, China}

\date{\today}

\begin{abstract}
We report the first complete calculation of the five heavy-to-light hadronic structure functions underlying semi-leptonic heavy-quark decays at next-to-next-to-next-to-leading order ($\mathcal{O}(\alpha_s^3)$) in perturbative QCD. This theoretical advance, achieved via an innovative hybrid computational strategy, enables precision predictions for triple differential decay rates. The results are essential for harnessing the potential of high-precision experiments at Belle II, BES III, and LHCb. Selected applications of this work include a state-of-the-art prediction for the inclusive $B \to X_u \ell \nu$ width, crucial for a percent-level determination of $|V_{ub}|$, and the first $\mathcal{O}(\alpha_s^3)$ results for lepton-energy moments in charm decays, vital for extracting $|V_{cs}|$ and $|V_{cd}|$. Our analysis also reveals significant higher-order corrections in the large-$q^2$ region of $b \to u$ transitions, offering new insights into the persistent tension between inclusive and exclusive $|V_{ub}|$ determinations.
\end{abstract}

\maketitle
\allowdisplaybreaks


\sect{Introduction}
The semi-leptonic~(sl.)~decays of heavy quarks or heavy-flavored hadrons constitute a unique natural laboratory for probing fundamental aspects of the Standard Model, including the CKM mechanism, the dynamics of QCD, and the determination of heavy-quark masses and decay widths.
The extraordinary experimental prospects~\cite{HFLAV:2022esi,HeavyFlavorAveragingGroupHFLAV:2024ctg,FCC:2018evy,CLICdp:2018esa,Belle-II:2022cgf,Ai:2024nmn} 
create a pressing need for equally or more precise theoretical predictions, to reduce theoretical uncertainties ideally to a level, e.g.~one third of the experimental ones, so that any persistent discrepancy can be unambiguously attributed to New Physics.
Improvements in various theoretical aspects of sl.~heavy-quark decays are thus much in need, among which the high-order perturbative QCD (pQCD) calculations are crucial for connecting the measured observables to the Standard-Model parameters and for reducing theoretical uncertainties.

A direct determination of $|V_{ub}|$ is feasible based on the analysis of charmless sl.~B~decays, either in the inclusive $\BXulv$ (i.e.~summing up all final-state hadronic d.o.f.s) or exclusive, e.g.~$\Bpilv$, channels, while the current results from the inclusive and exclusive determinations disagree significantly~\cite{HFLAV:2019otj,HFLAV:2022esi,ParticleDataGroup:2022pth}. 
The $|V_{ub}|$ extracted from inclusive $\BXulv$ also varies depending on the chosen precision observables and kinematic fiducial regions~\cite{Belle-II:2018jsg,HFLAV:2019otj}, 
suggesting that inconsistent experimental or theoretical inputs or underestimated systematic errors as well as New Physics are all among the possible reasons~\cite{Belle-II:2018jsg,Gambino:2020jvv}.
To achieve the unprecedented percent-level inclusive determination of $|V_{ub}|$ and resolve the long-standing tension with the exclusive values at Belle II~\cite{Belle-II:2018jsg,Belle-II:2022cgf}, precise measurements of as many independent \textit{differential} decay rates for $\BXulv$ as possible are the key ingredients~\cite{Bauer:2000xf,Bauer:2001rc,Belle:2003vfz,Belle:2021ymg}. 
To this end, higher-order pQCD corrections to the differential spectra of $\BXulv$, in particular the triple differential distribution, are among the essential theoretical inputs~\cite{Bosch:2004bt,Andersen:2005mj,Aglietti:2006yb,Gambino:2007rp,Gambino:2020jvv,Broggio:2026edk}. 
These results are required to apply necessary kinematic cuts to suppress backgrounds~\cite{Bauer:2000xf,Bauer:2001rc,Czarnecki:2001cz,Aglietti:2006yb,Gambino:2007rp}, in order to focus on the fiducial region more sensitive to the precision extraction of the non-perturbative parameters and the shape functions~\cite{Bigi:1993fe,Bigi:1993ex,Neubert:1993ch,Neubert:1993um,Bosch:2004bt,Bosch:2004th}, where unluckily the pQCD corrections are also enhanced and crucial~\cite{Andersen:2005mj,Aglietti:2006yb,Gambino:2007rp}. 
Moving to lower energy scales, despite the concern about the convergence of pQCD corrections for $\Dqlv$ because of the relatively large $\alpha_s(m_c)$, the higher-order results are clearly longed-for in the comprehensive study~\cite{King:2021xqp} of the inclusive sl.~decays of D~mesons. 
In this regard, the first simultaneous determination of $|V_{cs}|$ and $|V_{cd}|$ together with the non-perturbative theoretical parameters recently performed through a global fit~\cite{Shao:2025vhe,Shao:2025qwp} to the percent-level measurements of inclusive $\Dqlv$~\cite{CLEO:2009uah,BESIII:2021duu}, 
clearly demonstrates a promising prospect for incorporating high-order pQCD corrections for these processes.

As far as the above inclusive sl.~decays are concerned, the central underlying objects are the heavy-to-light hadronic structure functions $W_i$ that encode all relevant strong-interaction physics and underpin a vast array of precision observables, e.g.,~triple differential decay rates, lepton-energy and invariant-mass spectra, moments of distributions, and forward-backward asymmetries, etc. 
Despite huge theoretical efforts~\cite{Shifman:1984wx,Georgi:1990um,Bigi:1992su,Bigi:1993fe,Bigi:1993ex,Neubert:1992hb,Neubert:1993ch,Neubert:1993mb,Neubert:1993um,Mannel:1994pm,Bosch:2004bt,Andersen:2005mj,Gambino:2022dvu,Egner:2023kxw}, 
it is still very challenging to go beyond the next-to-leading order~\cite{Aquila:2005hq} (NLO,~$\Oals{}$) while maintaining the full kinematic dependence, 
and the ready-to-use $\mathcal{O}(\alpha_s^2)$ (N2LO) corrections remained only partial~\cite{Czarnecki:1994pu,Bosch:2004th,Lange:2005yw,Gambino:2006wk,Gambino:2007rp,Beneke:2008ei,Asatrian:2008uk,Brucherseifer:2013cu} until the very recent culmination~\cite{Broggio:2026edk} achieved for a subset of $\{W_i\}$ via a semi-analytic fit approach.  
In this work, by providing the first complete perturbative results for all five underlying $W_i$ and triple differential decay rates up to the unprecedented $\mathcal{O}(\alpha_s^3)$~(N3LO) accuracy, we directly address a critical and compelling bottleneck to harness the full potential of current~\cite{Belle:2021idw,Belle-II:2022evt,Belle:2021ymg,Belle:2023asa,BESIII:2021duu,BESIII:2024kvt} and future~\cite{Belle-II:2022cgf,Schwienhorst:2022yqu,Ai:2024nmn} precision experiments on inclusive sl.~weak decays of heavy quarks.

\sect{Method for calculating $W_i$}
The central quantity describing the inclusive sl.~decay of a heavy quark $Q$, with momentum $p$, into a collection of massless QCD partons with total momentum $p_X$, plus a pair of on-shell leptons with total momentum $k = p - p_X$ 
is the heavy-to-light hadronic tensor $\mathcal{W}_{Qq}^{\,\mu \nu}$. 
Its definition in perturbative QCD reads,%
{\small
\begin{align*} 
\mathcal{W}_{Qq}^{\,\mu \nu} = 
\frac{1}{\mathrm{N}} \, 
{\textstyle{\sum}} \hspace{-4mm} \int\limits_{X} 
d\Pi_X \, (2 \pi)^4 \delta^4(p - k - p_X) \, 
\langle p |\hat{J}_{Qq}^{\nu +}| p_X \rangle \langle p_X |\hat{J}_{Qq}^{\mu}| p  \rangle     
\end{align*}%
}
where $ d\Pi_X $ stands for the phase-space integration measure associated with the final-state QCD partons, and $J_{Qq}^{\mu}$ denotes the heavy-to-light weak-current operator $\bar{\psi}_Q\,\gamma^{\mu}(1-\gamma_5)/2\,\psi_q$.
The unpolarized inclusive $\HTensor$ depends on two independent momenta $\{p\,,k\}$, admitting the following Lorentz decomposition:%
\begin{align} \label{eq:HTffs}    
\mathcal{W}_{Qq}^{\,\mu \nu }(p, k) &= 
W_1 \,\big( -g^{\mu\nu}\,p^2 \big) \,+ \,
W_2 \, k^{\mu} k^{\nu} + \,
W_3 \, p^{\mu} p^{\nu}  \nonumber\\
&+ W_4 \,  \big(p^{\mu} k^{\nu} \, +\, k^{\mu} p^{\nu} \big) 
- W_5 \, i\epsilon^{\mu \nu \rho \sigma}\, p_\rho \, k_\sigma \,
\end{align}%
where the Levi-Civita tensor $\epsilon^{\mu \nu \rho \sigma}$ appears due to the chiral structure of the weak interaction. 
Each of the five form factors $W_i$, known as hadronic structure functions, depends on the Lorentz invariants $ k^2 = m_w^2,\, p\cdot k = m_Q\, E_w$ for given $p^2 = m_Q^2$, and receives both virtual and real-radiation type QCD corrections.
All loop and phase-space integrals involved are reduced by integration-by-parts (IBP) identities~\cite{Chetyrkin:1981qh,Anastasiou:2002yz} using {\tt Blade}~\cite{Guan:2024byi,Liu:2018dmc,Guan:2019bcx} combining with {\tt FiniteFlow}~\cite{Peraro:2019svx}. The resulting $\mathcal{O}(3000)$ bi-variate master integrals (MIs) are computed over the whole phase-space using an efficient hybrid approach that combines an efficient interpolation based on stratified Gauss-Kronrod points~\cite{Ehrich2000Mastroianni} with differential equations (DEs)~~\cite{Kotikov:1990kg,Gehrmann:1999as,Caffo:2008aw,Czakon:2008zk}.
Below, we sketch the main points behind this hybrid strategy, 
and more details and discussions can be found in the companion paper\cite{companionpaper}.

We first divide the domain in $y \equiv\, m_w^2/m_Q^2\, \in [0,\, 1]$ into 20 segments with equal length (sufficient for present application), and for each segment, a sample of points is chosen according to the 7-point Gauss-Kronrod rule. 
For each of these 140 points in $y$, the MIs are solved as univariate functions in $x \equiv E_w/m_Q$ between $\sqrt{y}$ and $(1 + y)/2$ using the DE method in terms of piecewise deeply-expanded series~(PSE) truncated up to 200 orders in $x$, but with the dimensional regulator~\cite{tHooft:1972tcz,Bollini:1972ui} $\varepsilon$ assigned with numerical values~\cite{Liu:2022mfb,Liu:2022chg} (see below).
For each requested point $\{x^{0}, y^{0}\}$, if $y^{0}$ coincides with any of the pre-sampled GK points, the values of MIs, or their linear combinations $W_i$, can be returned by directly evaluating the PSE solutions at very low computational cost. 
Otherwise, the values will be computed using an efficient interpolation based on the GK points selected within the very segment to which $y^{0}$ belongs; 
if the precision does not meet the requirement, more sample points from the neighboring segments may be incorporated, provided there are still enough significant digits left. 
Note that the function basis for the interpolation or linear-fit ansatz needs not be purely polynomials, but can contain inverse monomials and/or logarithmic factors in general. 
Additionally the dimensional regularization~\cite{tHooft:1972tcz,Bollini:1972ui}(DR) is employed to regularize the infrared and/or collinear (IR) divergences present in $W_i$ at the phase-space boundaries where $x$ reaches its maximum value for a given $y$, albeit with the dependence on the regulator $\varepsilon = (4-d)/2$ kept numerically~\cite{Liu:2022mfb,Liu:2022chg} in the PSE solutions. 
The phase-space integration of $\HTensor$ over the IR-dangerous regions in $x$ is done using its PSE with $\varepsilon$ assigned to numerical values $10^{-3} + n \times 10^{-4}$ for $n=0,~1$ (and a few more for inclusive quantities for intermediate cross-checks), in a way~\cite{companionpaper} compatible with the \textit{analytic} prescription of DR.
The fit regarding the $\varepsilon$-dependence is done only at the very end for the final finite physical objects of interest, which can be the IR-safe inclusive and differential decay widths, and this reduces the computational cost significantly. 
Armed with this hybrid strategy and highly efficient computational techniques, we have managed to obtain the high-precision semi-numerical results for all five $W_i$, and consequently $\HTensor$, in the whole phase-space in the two-dimensional plane $\{m_w^2,\, E_w\}$ up to $\mathcal{O}(\als^3)$ for the first time.

\begin{figure}[htbp]
  \begin{subfigure}[b]{0.20\textwidth}
    \centering
    \includegraphics[scale=0.40]{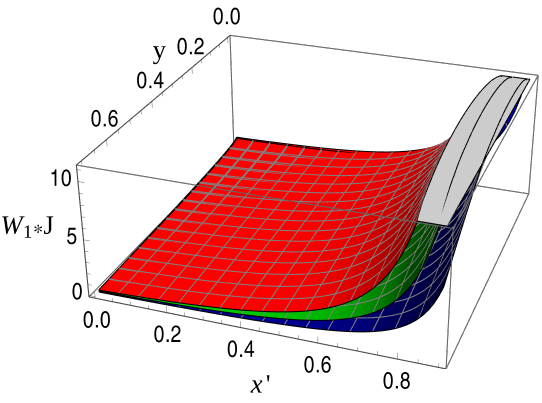}
  \end{subfigure}
  \begin{subfigure}[b]{0.20\textwidth}
    \centering
    \includegraphics[scale=0.40]{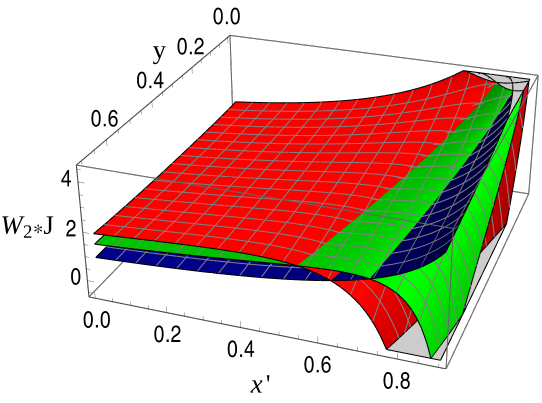}
  \end{subfigure}%
  \begin{subfigure}[b]{0.05\textwidth}
    \centering
    \includegraphics[scale=0.45]{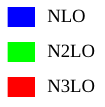}
  \end{subfigure}%

    \medskip

  \begin{subfigure}[b]{0.20\textwidth}
    \centering
    \includegraphics[scale=0.40]{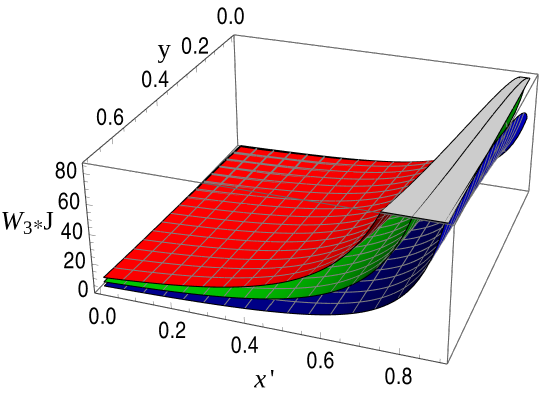}
  \end{subfigure}%
    \begin{subfigure}[b]{0.20\textwidth}
    \centering
    \includegraphics[scale=0.40]{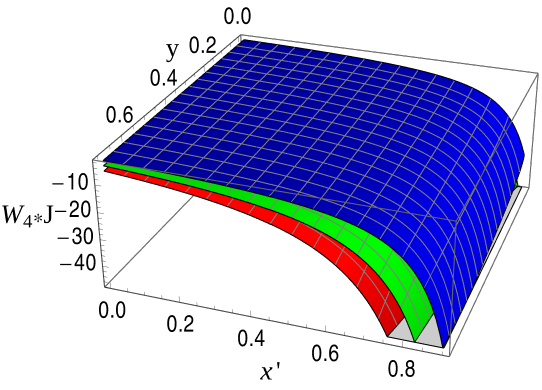}
  \end{subfigure}%
  \begin{subfigure}[b]{0.05\textwidth}
    \centering
    \includegraphics[scale=0.45]{figures/Legends_NLO_N2LO_N3LO.pdf}
  \end{subfigure}%

    \medskip

  \begin{subfigure}[b]{0.20\textwidth}
    \centering
    \includegraphics[scale=0.40]{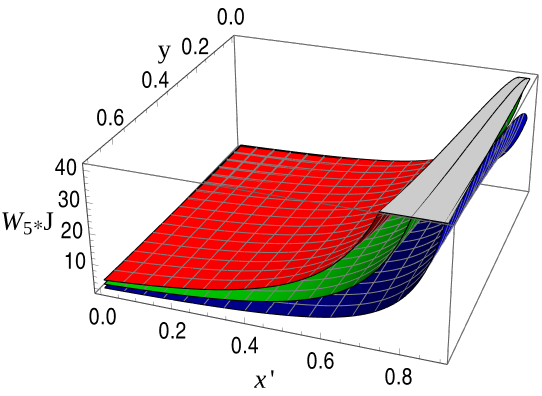}
  \end{subfigure}%
  \begin{subfigure}[b]{0.20\textwidth}
    \centering
    \includegraphics[scale=0.43]{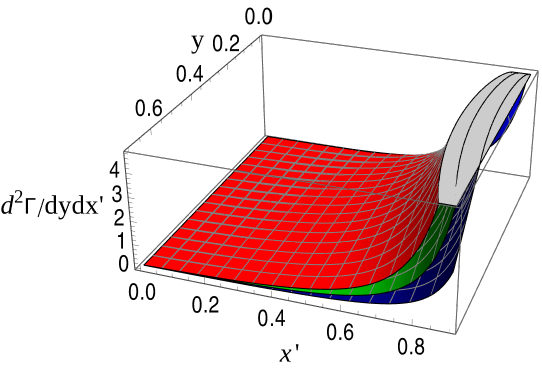}
  \end{subfigure}%
    \begin{subfigure}[b]{0.05\textwidth}
    \centering
    \includegraphics[scale=0.45]{figures/Legends_NLO_N2LO_N3LO.pdf}
  \end{subfigure}%
\caption{ {\small 
Top row: from left to right are the plots for the numerical results of $W_1\,, W_2$ in a regular phase-space region $R_2$ 
up to $\Oals{3}$. 
Middle row: from left to right are for $W_3\,, W_4$.
Bottom row: 
from left to right are for $W_5$ and the distribution~\eqref{eq:2Dparametric} plotted in the similar way.
}}
\label{fig:WsPlots}
\end{figure}

\sect{Results for $W_i$ and triple differential decay rates} 
Working with massless leptons, the triple differential decay rate of a heavy quark reads%
{\small
\begin{align} \label{eq:3Dparametric}
\frac{\mathrm{d}^3\,\Gamsl}{\mathcal{N}\, \mathrm{d} m_w^2\, \mathrm{d} E_w\, \mathrm{d} E_l} &= 
W_1\,  m_Q^2 m_w^2 + W_3\, m_Q^2 \big(2\,E_w\,E_l  - 2\, E_l^2 - m_w^2/2 \big) \nonumber\\
&+ W_5\, m_Q\, m_w^2 \big(E_w -2\, E_l \big)\,,
\end{align}
}
which involves three $W_i$.  
Since $W_i$ are independent of $E_l$, the dependence of $\mathrm{d}^3\,\Gamsl / (\mathcal{N}\, \mathrm{d} m_w^2\, \mathrm{d} E_w\, \mathrm{d} E_l)$ on $E_l$ takes a polynomial up to the second power. 
Once integrated over $E_l$, $W_5$ drops from the resulting distribution in $m_w$ and $E_w$. 
With all hadronic phase-space integrated in the definition~\eqref{eq:HTffs} of $W_i$, the remaining 6-fold phase-space integration measure can be further reduced to  
\begin{align*}
\int \mathrm{d}\, \mathrm{PS}_{L}
&= \frac{\pi^2}{ (2 \pi)^6 }\, 
\int_{0}^{m_Q^2} \mathrm{d} m_w^2 \, \int_{m_w}^{E_w^{\mathrm{max}}} \mathrm{d} E_w\, \int_{E_l^{\mathrm{min}}}^{E_l^{\mathrm{max}}} \mathrm{d} E_l
\end{align*}
with $E_w^{\mathrm{max}} \equiv (m_Q + m_w^2/m_Q)/2$, $E_l^{\mathrm{max}} \equiv (E_w + \sqrt{E_w^2 - m_w^2})/2 $ and $E_l^{\mathrm{min}} \equiv (E_w - \sqrt{E_w^2 - m_w^2})/2$, after eliminating the trivial d.o.f.s~irrelevant in unpolarized decays.  
We are now ready to present the interpolation plots for the numerical results of $W_i$ in a regular region $R_2$ of the two-dimensional phase-space in Fig.~\ref{fig:WsPlots}, up to $\mathcal{O}(\alpha_s^3)$ in a renormalization scheme where the heavy-quark mass $m_Q$ is renormalized in the on-shell (OS) scheme and $\als$ is $\MSbar$-renormalized. 
(The renormalization constants needed can be found in refs.~\cite{Larin:1993tp,vanRitbergen:1997va,Chetyrkin:1999pq,Czakon:2004bu,Melnikov:2000qh}.)
We have inserted $n_l = 4$ and $\alpha_s = 0.22$ for producing these benchmark plots, and set $m_Q=1$ as $W_i$ depend only on the dimensionless variable $x = E_w/m_Q$ and $y = m_w^2/m_Q^2$.%
The regular phase-space region $R_2$ is selected according to the kinematic constraint $y \in (0, 0.7)$ and $x \in (\sqrt{y}\,,\, 0.9\, x_{\mathrm{max}})$ with $x_{\mathrm{max}} = (1+y)/2$ to avoid the IR-singular edge located at $x = x_{\mathrm{max}}$.  
For better illustration, we have changed the coordinates of $R_2$ from the original $\{y,\,x\}$ into $\{y,\,x' = \frac{2\, (x - \sqrt{y})}{(1 - \sqrt{y})^2} \}$, which effectively stretches $R_2$ into a square as indicated by the horizontal plane in Fig.~\ref{fig:WsPlots}.
The Jacobian factor $J = (1 - \sqrt{y})^2/2$ generated by this change of variables is also included in the plots. 
In addition to the plots for five $W_i$, we have also plotted the double differential distribution obtained by integrating Eq.~\eqref{eq:3Dparametric} in $E_l$:%
\begin{align} \label{eq:2Dparametric}
\frac{\mathrm{d}^2\,\Gamsl}{\mathcal{N}\, \mathrm{d} y\, \mathrm{d} x} 
&=  
\int_{E^{\mathrm{min}}_l}^{E^{\mathrm{min}}_l}\, \frac{\mathrm{d}^3\,\Gamsl}{\mathcal{N}\, \mathrm{d} y\, \mathrm{d} x \, \mathrm{d}\, E_l}\, \mathrm{d}\, E_l \nonumber\\
&=
W_1  y \sqrt{x^2 - y} + \frac{1}{3} \big(x^2 - y \big)^{3/2} W_3
\end{align}%
in the right-most plot in Fig.~\ref{fig:WsPlots} in the same region $R_2$. 
Similar generalized ones with additional weight factors $E_l^N$ can be readily composed, the ``doubly-differential'' $N$-th moment of the lepton-energy spectrum~\cite{Gambino:2022dvu}.

\begin{figure}[htbp]
  \begin{subfigure}[b]{0.21\textwidth}
    \centering
    \includegraphics[scale=0.46]{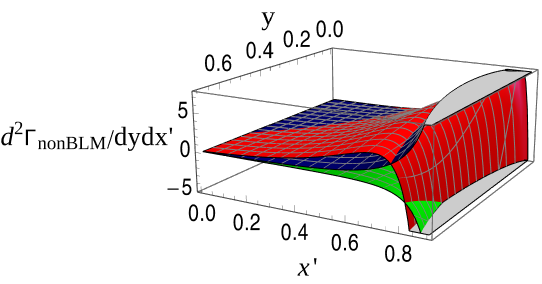}
  \end{subfigure}%
  \begin{subfigure}[b]{0.05\textwidth}
    \centering
    \includegraphics[scale=0.45]{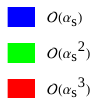}
  \end{subfigure}%
\begin{subfigure}[b]{0.21\textwidth}
    \centering
    \includegraphics[scale=0.46]{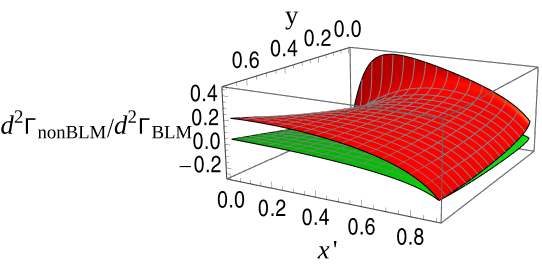}
  \end{subfigure}%
 \begin{subfigure}[b]{0.05\textwidth}
   \centering
   \includegraphics[scale=0.45]{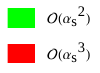}
  \end{subfigure}%
\caption{{\small
The left plot shows the perturbative coefficients after subtracting $\mathcal{O}(\alpha_s^{N+1} \beta_0^{N})$ terms with $N \geq 1$, 
with the $\Oals{1}$-coefficient included for reference. 
The right plot shows the ratios of the non-BLM to the BLM-type contribution at, respectively, $\Oals{2}$ and $\Oals{3}$.}}
\label{fig:BLMPlots}
\end{figure}

A few comments are in order.
First of all,  all contributions covered in the regular region in Fig.~\ref{fig:WsPlots} must have at least one real QCD-parton with non-vanishing energy and non-collinear momentum, and thus begin at $\mathcal{O}(\alpha_s)$.
Secondly, the pQCD corrections all become larger as $y$ increases, and rise rapidly as $x$ (or its rescaled counterpart $x'$) gets close to its maximum value. 
This is expected as $W_i$ shall not be finite at $x = x_{\mathrm{max}}$. 
Due to the relatively large $\alpha_s$ at the energy scales of $b$-~and $c$-quark decay, the convergence behavior of the first few perturbative terms is already noticeably affected by the leading IR-renormalon related to the quark pole mass~\cite{Bigi:1994em,Beneke:1994sw,Beneke:1994rs,Smith:1996xz}. 
Related to this, for $W_i$ and the differential decay rates~\eqref{eq:2Dparametric}, the $\mathcal{O}(\alpha_s^2 \beta_0)$ BLM-type~\cite{Brodsky:1982gc} corrections are expected to constitute the major part of the $\Oals{2}$ corrections~\cite{Andersen:2005mj,Aquila:2005hq,Gambino:2006wk,Gambino:2007rp,Brucherseifer:2013cu} in the OS scheme, although the exact ratios depend on the kinematic regions. 
Here we confirm this expectation at $\Oals{2}$ and extend further the analysis to $\Oals{3}$ as shown in Fig.~\ref{fig:BLMPlots}. 
We define the BLM-part $\mathrm{d}^2\,\Gamsl_{\mathrm{BLM}} $ by collecting all $\mathcal{O}(\alpha_s^{N+1} \beta_0^{N})$ corrections for $N \geq 0$, thus the so-defined non-BLM part starts only from $\Oals{2}$. 
From the ratio plot in Fig.~\ref{fig:BLMPlots} we confirm, up to $\Oals{3}$, that for a given $q^2 = y\, m_Q^2$ in most of the covered regular region, the larger $E_w$ (i.e.~$x'$), the BLM-type contributions dominate more over the non-BLM-type contributions.  
After subtracting the $\mathcal{O}(\alpha_s^{N+1} \beta_0^{N})$ terms with $N \geq 1$, the remaining pQCD contributions exhibit a much better convergence behavior up to $\Oals{3}$ in the regular region as indicated by the left plot. 
We thus anticipate that merging our high-order results  with the BLM-resummation~\cite{Gambino:2006wk} will lead to further improved perturbative predictions highly valuable for phenomenological studies for $\bulv$ and $\cqlv$, which we leave for near-future work. 

\sect{Applications to selected precision observables}
Due to different energy scales and physical observables involved in the applications to the sl.~weak decays of $b$-~and $c$-quark, the best practices to treat the perturbative QCD corrections vary, and it is thus more convenient to discuss different representative applications separately.

{\it 1.~Results for $\GamBulv$.} 
The application of the Operator-Product-Expansion (OPE) to the inclusive charmless sl.~B~decays $\BXulv$ yields the formula~\cite{Bigi:1992su,Bigi:1993fe,Neubert:1993ch,Neubert:1993mb,Uraltsev:1999rr} for $\Gamma\big(\BXulv\big)$: 
{\small
\begin{align} \label{eq:gammaB2Xulv}
\Gamma_0(m_b) \big[
\mathbf{C}_{\mathrm{p}}\,\big(
1 - \frac{\mu_\pi^2 - \mu_G^2}{2\, m_b^2}
\big) - 2 \frac{\mu_G^2}{m_b^2} + \mathcal{O}\big(\Lambda_{\mathrm{QCD}}^3/m_b^3\big)\big]\,,    
\end{align}
}
where $\mathbf{C}_{\mathrm{p}}$ denotes the perturbative QCD correction factor to the free $b$-quark decay width $\GamBulv \equiv \Gamma_0(m_b)\,\mathbf{C}_{\mathrm{p}}$, 
and the $1/m_b^2$-suppressed non-perturbative contributions $\mu_{\pi}^2$ and $\mu_{G}^2$ are, respectively, the expectation values of the kinetic and chromomagnetic operators~\cite{Bigi:1992su,Bigi:1993fe,Neubert:1993mb}. 
$\Gamma_0 \equiv \frac{G^2_F\,  |V_{ub}|^2\, m_b^5\, A_{\mathrm{ew}}}{192 \pi^3}$ encodes the known electroweak K-factor $A_{\mathrm{ew}} = 1.01435$~\cite{Sirlin:1981ie,Fael:2024fkt}.
Despite the considerable improvement in the behavior of the perturbative series for $\GamBulv$ in $\MSbar$ mass over the OS scheme, we find that its conventional scale uncertainty at N3LO is still as large as $[+6\%, -8\%]$~\cite{companionpaper}.
Further improvement is achieved by rewriting $\GamBulv$ using the kinetic mass~$m_b^{\mathrm{kin}}[1] \equiv m_b^{\mathrm{kin}}(\mu_c=1\mathrm{GeV}) = 4.554 \pm 0.018$~GeV\cite{HFLAV:2019otj}, leading to:%
\begin{align}
\label{eq:Gkin}
\Gamma_0(m_b^{\mathrm{kin}}[1]) \, 
\big(
1\, -\, 0.027108 \,+\, 0.02509 \,+\, 0.02182
\big)
\end{align}%
at the wittily chosen $\mu=m_b^{\mathrm{kin}}[1]/2$ for the 4-flavor $\alpha_s(\mu)$~\cite{Bordone:2021oof}.  
Our new full results at $\mathcal{O}(\als^2)$ and $\mathcal{O}(\als^3)$ in the kinetic mass scheme increasing the N2LO result~\cite{Uraltsev:1999rr} by about $2\%$. 
Incorporating the $\mathcal{O}(1/m_b^2)$ non-perturbative corrections in Eq.~\eqref{eq:gammaB2Xulv} given by $\mu_{\pi}^2 = 0.477 \pm 0.056 \mathrm{\,GeV^2}$~and $\mu_{G}^2 = 0.306 \pm 0.050 \mathrm{\,GeV^2}$~\cite{Bordone:2021oof}, we finally obtain the most precise theoretical prediction for $\Gamma(\BXulv)$: 
{\small
\begin{align*}
\frac{|V_{ub}|^2}{|3.82\times 10^{-3}|^2}\,\big( 6.53 \,\pm 0.12 \, \pm 0.13\, \pm 0.03\, \big) \times 10^{-16}\,\text{GeV}\,,    
\end{align*}
}
where the first and third errors are related to the fit errors of the input $m_b^{\mathrm{kin}}[1]$, $\mu_{\pi}^2$ and $\mu_{G}^2$; 
the second error is assigned from the residual scale uncertainty estimated conservatively by $\mu \in [m_b^{\mathrm{kin}}[1]/2\,, 2\,m_b^{\mathrm{kin}}[1] ]$.

\begin{figure}[htbp]
\includegraphics[scale=0.5]{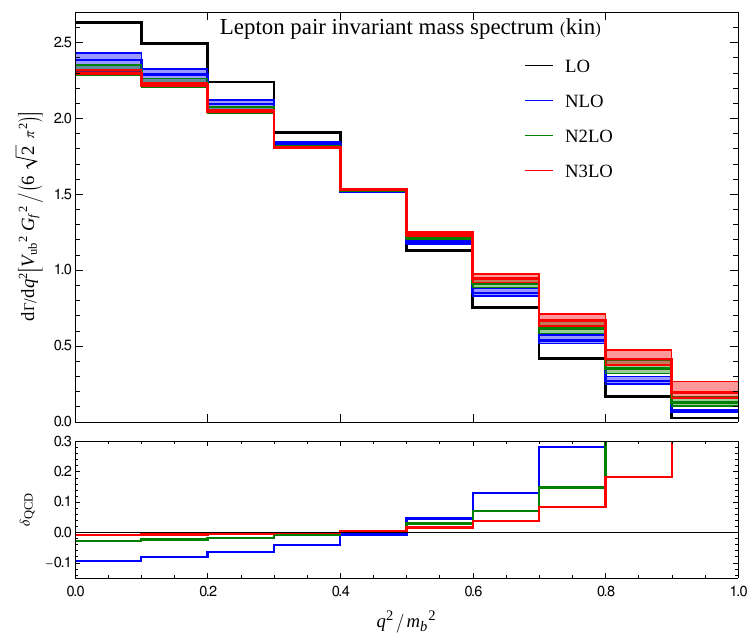}
\caption{{\small
The perturbative results for the histogrammed $q^2$-spectrum determined in the kinetic mass scheme at different orders in $\alpha_s$ with the conventional scale variation $\mu \in [\mbkin/2, \mbkin]$. 
}}
\label{fig:b2ulv_dGamdq2_Kin}
\end{figure}

{\it 2.~Results for $q^2$-spectrum in $\bulv$.} 
The leptonic $q^2$-spectrum in $\BXulv$ was proposed~\cite{Bauer:2000xf,Bauer:2001rc} as one of the precision observables for inclusive determination of $|V_{ub}|$, and has been measured at Belle II~\cite{Belle:2003vfz,Belle:2021ymg}, where a lower cut on $q^2$ is necessary to suppress the $\BXclv$ background. 
Here we are only concerned with pQCD corrections defined initially in the OS scheme:  
{\small
\begin{align*}
\int^{\mbos^2}_{0} \frac{\mathrm{d} \GamBulv }{\mathrm{d} q^2}  \mathrm{d} q^2 
= \int^{\mbos^2}_{0} \big( f_0 + \als f_1 + \als^2 f_2+ \als^3 f_3  \big) \mathrm{d} q^2\,.    
\end{align*}
}
Our full result for $q^2$-spectrum in the OS scheme is cross-checked against the expanded $\mathcal{O}(\alpha_s^2)$ result~\cite{Czarnecki:2001cz} valid in the large-$q^2$ region, and the partial results~\cite{Chen:2023osm,Chen:2023dsi,Fael:2023tcv}. 
The OS scheme is clearly not the best choice here.
However, changing to other quark-mass schemes, consistently and thoroughly, for the $q^2$-spectrum at the differential level entails additional subtlety as compared to its inclusive moments. 
This may be illustrated by the perturbative re-expansion of the $m$-dependence of the prototype integral $\int_{0}^{m}\, f(\alpha_s, m;\, x)\, \mathrm{d} x$:  
{\small
\begin{align}\label{eq:integralpowexp}
&\int_{0}^{m_0}\, f(\alpha_s,m_0 + \delta_m;\, x)\, \mathrm{d} x  + 
\int_{m_0}^{m}\, f(\alpha_s, m_0 + \delta_m ;\, x) \, \mathrm{d} x  \nonumber\\ 
&= \int_{0}^{m_0}\, 
\sum_{n=0}^{\infty} \frac{1}{n!}\, \frac{\partial^n\, f(\alpha_s\,,\, m \,;\, x)}{\partial\, m^n}\Big|_{m=m_0} \, \big(\delta m\big)^{n} \, \mathrm{d} x  \nonumber\\ 
&+ 
\sum_{k=0}^{\infty} \frac{1}{(k+1)!}\frac{\partial^k f(\alpha_s\,,\, m_0 + \delta_m ;\, x)}{\partial\, x^k}\Big|_{x=m_0} \big(\delta m \big)^{k+1}\,,
\end{align}
}
under the mass transformation $m  = m_0 + \delta m $ with $\delta m $ starting from $\mathcal{O}(\alpha_s)$.
The first part, with $x$ integrated over $[0,\, m_0]$, is usually expected, where all re-expansions involve only the local $m$-dependence of the regular $f$.
Further series expansion of the second part eventually involves only $f(\alpha_s\,,\, m_0\,;\, x=m_0)$ and its partial derivatives. 
Specialized to $b \rightarrow u \ell \bar{\nu}_\ell$ at hand, our explicit computations show that these \textit{boundary-effect} terms for the $q^2$-spectrum become non-vanishing but only starting from $\Oals{3}$!~(See the companion paper~\cite{companionpaper} for details.)
Incorporating these terms is necessary to \textit{exactly} preserve the integrity of integrated moments of the perturbatively re-expanded $q^2$-spectrum. 
However this can only be meaningfully done in terms of the \textit{histogrammed} $q^2$-spectrum, where these boundary-effect terms are attributed to the last bin covering phase-space boundaries. 
In the limit of an infinitesimal binning, the non-zero boundary-effect terms will manifest themselves in the form of a Dirac-$\delta$ modification to the distribution around $x=m_0$.
Our novel numerical results for the histogrammed $q^2$-spectrum determined in the kinetic mass scheme are shown in Fig.~\ref{fig:b2ulv_dGamdq2_Kin}, where pQCD corrections exhibit decent convergence and regular behaviors.
The non-vanishing $\mathcal{O}(\als^3)$ boundary-effect term has been included in the right-most bin at $q^2/ \mbkin^2 = 1$ and affects the bin-value by $3 \sim 5\%$ (depending on the choice of scales). 
It is particularly interesting to note that the sign of $\delta_{\mathrm{QCD}}^{(N)} = \alpha_s^N\, f_N / \sum_{i=0}^{N} \alpha_s^i\, f_i$ in Fig.~\ref{fig:b2ulv_dGamdq2_Kin} flips when moving from the low-$q^2$ to the high-$q^2$ region at each order in $\alpha_s$, crossing zero at about the middle region of the $q^2$-domain. 
This particular crossing pattern thus offers us an illuminating perspective to appreciate the previously puzzling feature exhibited in the perturbative corrections to $\GamBulv$ in~Eq.~\eqref{eq:Gkin}: the sizes of these corrections, although small, do not decrease much as one goes to higher orders. 
On the other hand, when restricted to the large-$q^2$ region, say $q^2 > (m_b-m_c)^2$~\cite{Bauer:2000xf,Bauer:2001rc}, the $\mathcal{O}(\als^3)$ corrections are quite sizable, clearly shown in Fig.~\ref{fig:b2ulv_dGamdq2_Kin}. 
More specifically, the QCD correction factor in the last 5 bins in total reads 
$1 + 0.1647 + 0.1264 + 0.09519 = 1.386$ with the conventional scale uncertainty $[-4 \%, 6\%]$, and the impact of the $\Oals{3}$ correction is thus significantly larger than for the inclusive moments in the kinetic mass scheme.  
(We have observed similar features in the results derived in the 1S scheme~\cite{companionpaper}.)
Given the important role of this region in extracting $|V_{ub}|$, it would be interesting to investigate the potential impact of our findings in clarifying the puzzling discrepancy between the inclusive and exclusive determination of $|V_{ub}|$ in future fully fledged analyses.

{\it 3.~Results for lepton-energy moments in $\cqlv$.} 
As the last highlight of our selected applications, we present the first N3LO perturbative corrections to the  inclusive electron-energy moments (EEM) in $\cqlv$, whose hadronic counterparts in $D_s \rightarrow X\,\ell \bar{\nu}_{\ell}$ are important observables sensitive to $|V_{cq}|$ and have been measured to 
percent-level precision at BES III~\cite{BESIII:2021duu,DeSantis:2025qbb}.
To be definite, we consider the perturbative (un-normalized) EEMs for $c \rightarrow q\,\ell \bar{\nu}_{\ell}$: 
$\int^{E_e^{\mathrm{max}}}_{0} 
E_e^{N} \, \frac{\mathrm{d} \GamBulv }{\mathrm{d}\, E_e} \, \mathrm{d}\, E_e\,
\equiv \frac{G_F^2\, |V_{cq}|^2}{192 \, \pi^3} \, \langle E_e^{N} \rangle\big|_{unn.}$  
Using the 1S mass of $c$ quark $\mcOs = 1.55$~GeV~\cite{ParticleDataGroup:2024cfk} and $\als(\mu=\mcOs) = 0.338$, our numerical results for the perturbative EEMs at $\mu = \mcOs$ read:%
{\small \begin{align*}
\langle E_e^{0} \rangle \big|_{unn.} &= 8.9466 - 1.1871 ~\epsilon - 0.4440 ~\epsilon^2 - 0.0609 ~\epsilon^3 + \mathcal{O}(\epsilon^4) \nonumber\\
\langle E_e^{1} \rangle \big|_{unn.} &= 4.1602 - 0.4994  ~\epsilon - 0.1307 ~\epsilon^2 + 0.0861 ~\epsilon^3 + \mathcal{O}(\epsilon^4) \nonumber\\
\langle E_e^{2} \rangle \big|_{unn.} &= 2.1494 - 0.2311 ~\epsilon - 0.02498 ~\epsilon^2 + 0.1107 ~\epsilon^3 + \mathcal{O}(\epsilon^4) \,,
\end{align*}
}%
where the formal power-expansion parameter $\epsilon$ in the 1S-expansion prescription is kept to track the order of perturbation theory~\cite{Hoang:1998ng,Hoang:1999zc} (that will be set 1 to resume the final result). 
The 0-th EEM is precisely the inclusive decay width for $c \rightarrow q\,\ell \bar{\nu}_{\ell}$, and 
it is a bit surprising to observe a decent convergence behavior up to $\mathcal{O}(\epsilon^3)$.
It is quite encouraging that the above results up to $\mathcal{O}(\epsilon^2)$ have been successfully employed and proven instrumental in improving the precision of the first simultaneous determination~\cite{Shao:2025vhe,Shao:2025qwp} of $|V_{cs}|$ and $|V_{cd}|$ together with relevant non-perturbative parameters, and we look forward to seeing the impact of our higher-order results.

\sect{Summary}
We have accomplished the first complete calculation of all five heavy-to-light structure functions $W_i$ underlying the triple differential semi-leptonic decays of heavy quarks up to $\mathcal{O}(\alpha_s^3)$, representing a major advancement in improving the pertubative precision of theoretical predictions for these processes.
Among the highlights of the selected applications, we present the state-of-the-art theoretical prediction $\Gamma(\BXulv) =  \frac{|V_{ub}|^2}{|3.82\times 10^{-3}|^2}\,\big( 6.53 \,\pm 0.12 \, \pm 0.13\, \pm 0.03\, \big) \times 10^{-16}\,\text{GeV}\,$ derived in the kinetic mass scheme, which serves as an indispensable theoretical input for achieving the percent-level inclusive measurement of $|V_{ub}|$ at Belle II.
In addition, the sign-flip pattern of pQCD corrections to the $q^2$-spectrum offers us an illuminating perspective to appreciate a puzzling feature exhibited in the inclusive $\Gamma(\BXulv)$. 
On the other hand, it indicates sizable $\mathcal{O}(\als^3)$ corrections when restricted to the large-$q^2$ region  
observed in both kinetic and 1S mass schemes used in the past analyses. 
Given the important role of the region with large $q^2$ in measuring $|V_{ub}|$, it would be interesting to further investigate the potential impact of our findings.  
We highlight also a novel interesting point in the consistent perturbative reformulation of the differential $q^2$-spectrum from the pole-mass to other mass schemes: certain boundary-effect terms are identified that firstly become non-vanishing at $\mathcal{O}(\alpha_s^3)$ for $\bulv$; 
their incorporation is essential to preserve the integrated moments of the perturbatively re-expanded $q^2$-spectrum but necessitates histogramming from $\mathcal{O}(\alpha_s^3)$ onward even within pure perturbation theory.

Looking forward, the hybrid strategy employed to compute $W_i$, which combines an efficient linear interpolation (using a suitable function basis that needs not be polynomial) based on stratified Gauss-Kronrod points in one dimension and the differential equations in the other degree(s) of freedom, further armed with reduced numerical $\varepsilon$-dependence, can be applied, e.g.~to $\BXclv$, to take into account the final-state quark-mass effects.  
Merging with the BLM-resummed results will yield more accurate perturbative predictions, which are highly valuable for phenomenological studies.
The work presented for both the inclusive and differential semi-leptonic decay rates, and those soon becoming available with our approach, at the unprecedented accuracy of $\mathcal{O}(\alpha_s^3)$ will help to achieve the physical goal of percent-level inclusive measurments of CKM matrix elements (e.g.~$|V_{ud}|$, $|V_{cs}|$ and $|V_{cd}|$) and non-perturbative dynamical parameters at Belle II, BES III and LHCb, advancing the precision frontier for these important processes to a new level. 

\begin{acknowledgments}
\sect{Acknowledgments}
The work of L.C. was supported by the Natural Science Foundation of China under contract No.~12205171, No.~12235008 and No.~12321005, and Department of Science and Technology of Shandong province No.~tsqn202312052 and~2024HWYQ-005.
X.C. was supported by the Swiss National Science Foundation (SNF) under contract 200020\_219367 and the UZH Postdoc Grant, grant no. [FK-25-104].
X.G. was supported by the United States Department of Energy, Contract DE-AC02-76SF00515. 
Y.Q.M. was supported in part by the National Natural Science Foundation of China under contract No.~12325503.

\textbf{Note Added: }
When finalizing this paper, we noticed that the other group~\cite{Broggio:2026edk} had just completed a semi-analytic extraction of the hadronic structure functions $W_1, W_3, W_5$ (relevant for $\bulv$ with massless leptons) at $\Oals{2}$. In the region with low $q^2 < 0.5\, m_Q^2$ and relatively large $q\cdot p /m_Q$, where the numerical errors in Ref.~\cite{Broggio:2026edk} are small, we find very good agreement for the pure $\Oals{2}$ corrections to $W_1, W_3, W_5$, with the relative difference below 1\%.

\end{acknowledgments}

\bibliographystyle{utphysMa}
\bibliography{H2Ldecay}

\end{document}